# Skyrmion mediated voltage controlled switching of ferromagnets for reliable and energy efficient 2-terminal memory

*Dhritiman Bhattacharya[1] and Jayasimha Atulasimha[1, 2] ‏*

[1]Department of Mechanical and Nuclear Engineering

[2]Department of Electrical and Computer Engineering

Virginia Commonwealth University, Richmond, VA 23284, USA.

* Corresponding author: jatulasimha@vcu.edu

We propose a two terminal nanomagnetic memory element based on magnetization reversal of a perpendicularly magnetized nanomagnet employing a unipolar voltage pulse that modifies the perpendicular anisotropy of the system. Our work demonstrates that the presence of Dzyaloshinskii-Moriya Interaction (DMI) can create alternative route for magnetization reversal that obviates the need for utilizing precessional magnetization dynamics as well as a bias magnetic field that are employed in traditional voltage control of magnetic anisotropy (VCMA) based switching of perpendicular magnetization. We show with extensive micromagnetic simulation, in the presence of thermal noise, that the proposed skyrmion mediated VCMA switching mechanism is robust at room temperature leading to extremely low error switching while also being potentially 1-2 orders of magnitude more energy efficient than state of the art spin transfer torque (STT) based switching.

Static power dissipation in the CMOS devices is a major drawback and impedes downscaling of these devices [1]. Static power dissipation can be eliminated by combining a non-volatile memory with the volatile logic circuits and continued scaling with improved energy efficiency can be realized. Nanomagnetic memory devices, due to their inherent non-volatility and compatibility with standard CMOS process, are considered to be the one of the key contenders [2], [3]. A typical nanomagnetic memory element is built on the magnetic tunnel junction (MTJ) shown in Fig 1a. The free layer of the MTJ is switched between two stable orientations which changes the tunnel magnetoresistance (TMR)[4]–[7] and thus allows reading out of the memory bit. Various mechanisms for switching the free layer magnetization are being investigated to implement energy efficient, scalable and high endurance memory elements. Some techniques that have potential to attain these goals include spin current induced switching[8]–[12] and voltage controlled switching of nanomagnets[13]–[21]. Among spin current based switching strategies, spin transfer torque (STT) induced switching[8]–[10] requires a large current which subsequently translates into a large energy dissipation (~100 fJ/bit). This is at least three orders of magnitude larger than the energy dissipation in CMOS (~100 aJ/bit)[22]. This large write energy dissipation negates the advantage of non-volatility offered by nanomagnetic memory. An alternative method of switching the magnetization is by the generation of spin current using giant spin hall effect (SHE)/spin orbit torque (SOT) [11], [12]. However, three terminal device structure and requirement of high resistance ultrathin metal interconnect limits scaling of these devices[23]. Electric field induced magnetization reversal, on the other hand, promises to be highly energy efficient while also being amenable to implement two-terminal memory device elements. Several studies examined the feasibility of ultra-low power memory devices that exploits voltage induced strain generation in a magnetostrictive material[19]–[21] or voltage controlled modulation of perpendicular anisotropy in a ferromagnet/oxide interface (VCMA)[13]–[18].

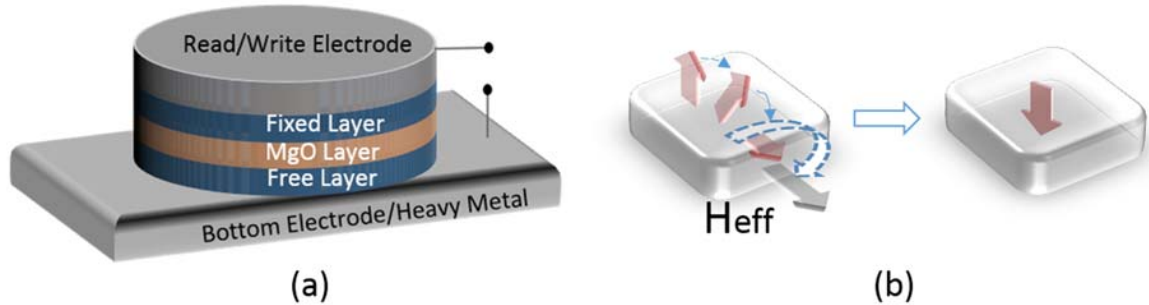

Figure 1. (a) Simple schematic representation of a magnetic tunnel junction, (b) precessional switching of the magnetization of the free layer in the presence of an in plane bias magnetic field

Recently, reduction of write energy down to 6fJ has been shown experimentally using VCMA in perpendicular MTJs[17]. Perpendicular nanomagnets having two stable magnetization orientations that point out of plane (±z direction) are the preferred candidate for the free layer of a magnetic tunnel junction due to better scalability than planar nanomagnets[24], [25]. Perpendicular anisotropy originates at a ferromagnet/oxide interface from the overlap between oxygen's $p_z$ and ferromagnet's hybridized $d_{xz}$ orbital [26]. Electron density at the interface can be altered by employing a voltage pulse which consequently changes the perpendicular anisotropy present in the system [27]. This phenomenon is called voltage control of magnetic anisotropy (VCMA). Evidently, application of a voltage pulse can only shift the easy axis from perpendicular to in plane direction. Therefore, the magnetization only experiences a 90° rotation and has an equal probability of ultimately settling in any of the two bistable orientations when the voltage pulse is withdrawn. An in-plane magnetic field is utilized to achieve complete magnetization reversal (i.e. 180° rotation. which is desired to maintain a high magnetoresistance ratio as well as to ensure nonvolatile operation). Upon application of a voltage pulse, the magnetization rotates to the planar orientation while also precessing about the magnetic field (Fig 1b). Complete reversal can be achieved by designing the magnetic field strength and the voltage pulse duration appropriately. In this paper, we show that ferromagnetic reversal in a system with Dzyaloshinskii-Moriya Interaction (DMI)[28], [29] obviates the need of using this magnetic field or precessional magnetization dynamics. We also show that such skyrmion mediated reversal is robust to room temperature thermal noise.

Dzyaloshinskii-Moriya Interaction (DMI), given by $H_{DMI} = -\vec{D_{12}}.(\vec{S_1} \times \vec{S_2})$ (where $D_{12}$ is the DMI vector, $S_1$ and $S_2$ are two neighboring atomic spins), can be significant at the interface of a ferromagnet and a heavy metal of high spin orbit coupling. This promotes canting between the magnetization of two neighboring spins and could lead to the stabilization of spiral spin orientation such as magnetic skyrmions[30]–[32]. Magnetic skyrmions are beneficial for implementing racetrack like memory devices as they require a very small current to move[33]–[35]. Fixed skyrmion based memory devices are also theoretically demonstrated where two bistable orientations are core-up and core-down skyrmions, and the switching agent can be magnetic field[36], [37], electric field[38]–[40], spin current[41], [42] or a combination of VCMA and spin current[43]. Presence of DMI imparts detrimental effect as it increases the required current density in STT induced ferromagnetic reversal [44], [45]. However, we have shown core reversal of a skyrmion(that is stabilized by DMI) can be more energy efficient than ferromagnetic reversal when VCMA is used in conjunction to spin current[43]. Furthermore, we showed, switching between two ferromagnetic states is possible in a system with DMI [38] by sequential increase and decrease of perpendicular anisotropy (i.e. applying a bipolar voltage pulse).

In this study, we theoretically demonstrate an energy efficient (sub-femtojoule), scalable (100 nm), fast (sub-nanosecond) and error free non-volatile memory element driven solely by a unipolar voltage pulse (i.e without any external magnetic field). Write error rate (WER) in the conventional VCMA induced switching strongly depends on the voltage pulse duration. Peak switching probability is attained when pulse width is set to half period of magnetization precession. However, experimental studies achieved WER of only $4\times 10^{-3}$[46], possible due to incoherent magnetization states during the switching process that result in substantially higher error rates than predicted by single domain simulations of magnetization dynamics. We note that the tolerable error limit for memory application can be $10^{-15}$ nessisating an iterative approach to lower the WER. For example WER of $10^{-9}$ is achieved with 4 iterations while 10 iterations were needed to improve the WER to $10^{-17}$[47]. This technique obviously consumes more energy and increases write time substantially. We evaluated the performance of our proposed method under room temperature thermal perturbation. Reliable performance with $<10^{-4}$ switching error in the pulse width range of 0.47 ns-0.58 ns was found. We note that we could only run 10,000 simulations (as solving the magnetization dynamics with thermal noise takes a lot of computational time) and saw no switching error. Hence we only claim $<10^{-4}$ switching error (which is an order magnitude better than other incoherent switching schemes in the presence of thermal noise [46]), while our scheme could be even more robust to switching in the presence of thermal noise. This method therefore has the potential to eliminate or at least reduce the iterative write process which will consequently lower the write time and energy dissipation.

We simulated the free layer of the MTJ structure in micromagnetic framework MuMax3[48]. Details of the micromagnetic simulation is discussed in the methods section. The free layer was chosen to be nanodisk of 100 nm diameter and 1 nm thickness. The material parameters used in the simulations are: saturation magnetization $M_s$=1.3×$10^6$ A/m, perpendicular anisotropy $K_{u1} = 1.1$ MJ/$m^3$, exchange stiffness $A_{ex}$=25 pJ/m, Gilbert damping coefficient α=0.01 and DMI parameter D=1.2 mJ/$m^2$. Effective perpendicular anisotropy, $K_{eff} = K_{u1} - \frac{1}{2}\mu_0 M_s^2$=4.5×$10^4$ J/$m^3$ without considering barrier reduction due to DMI. DMI value is less than the critical value ($D_{crit} = \frac{4}{\pi}\sqrt{AK_{eff}}$ =1.35 mJ/$m^2$) needed to form a skyrmion at this PMA. Hence, (quasi)ferromagnetic state emerges as the only stable state where the spins at the periphery slightly tilt towards the x-y plane.

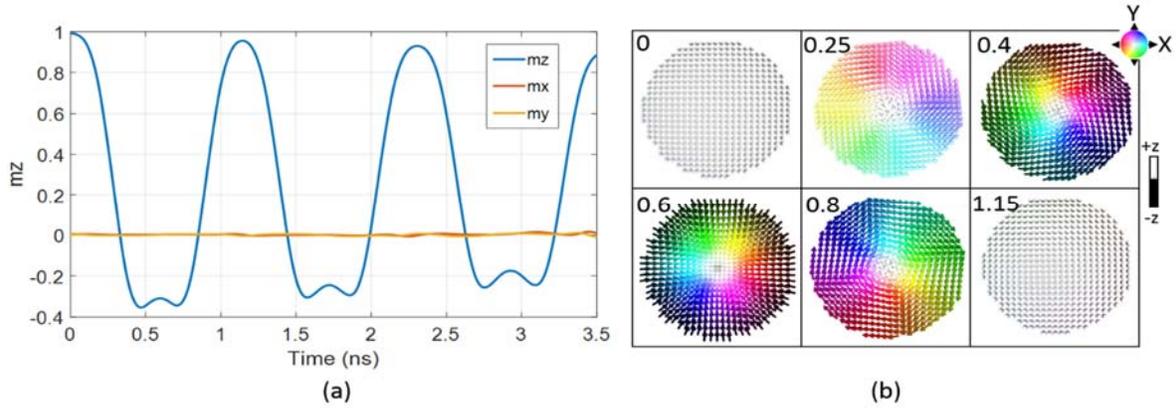

Figure 2: (a) Change in magnetization due to reduction in PMA, (b) Magnetization states visited during the first cycle of breathing mode, top left corner shows time in nanosecond.

Next, we study the behavior of the nanomagnet when PMA in the system is reduced through VCMA from 1.1 MJ/m$^3$ to 0.94 MJ/m$^3$. The magnetization dynamics is shown in Fig 2(a). We take a quasi-FM state with spins tilting slightly at the disk edge (Fig. 2b, 0 ns) as our initial state found by relaxation of an upward ferromagnetic state for 5 ns. When the PMA in the system is reduced, peripheral spins start to tilt more in the planar direction and undergo 180° rotation. The amount of rotation is smaller s for the spins that are closer to the core and spins at the core do not budge from their initial position. Therefore, we find a state where the spins at the core point up (+z) and the spins at the periphery are tilted downward (-z) and a core-up skyrmion is formed (Fig. 2b, 0.6 ns). Once the skyrmion is formed, breathing mode is excited due to which the skyrmion core expands and shrinks. Hence, the magnetization state of the system alternatively visits the skyrmionic and the quasi ferromagnetic state (Fig. 2b).

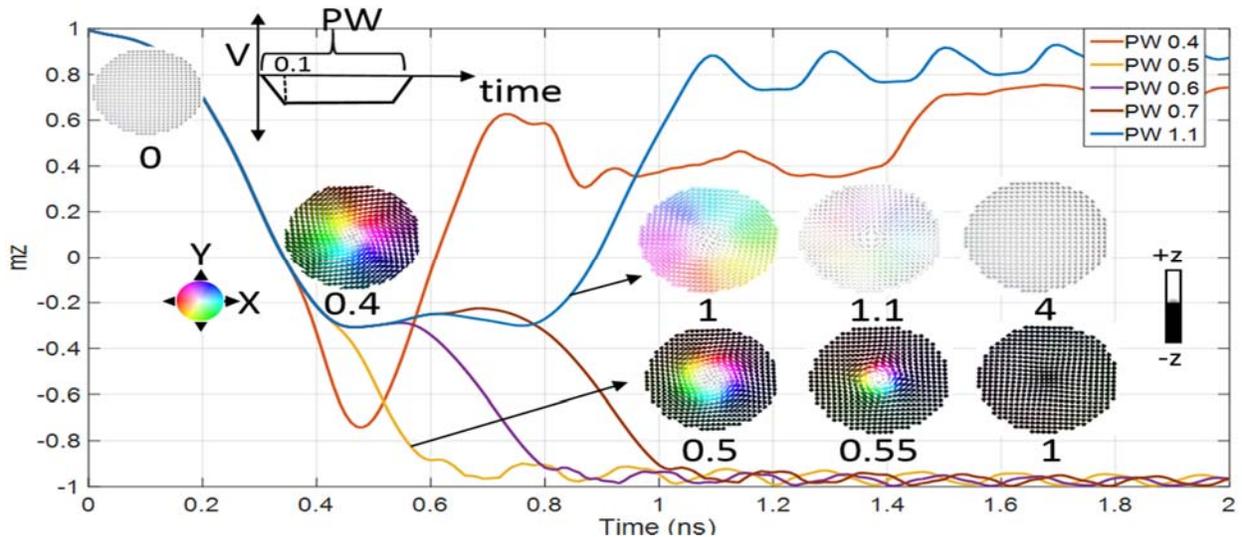

Figure 3: Magnetization dynamics in response to voltage pulse of different pulse width. The voltage pulse is shown on the top left corner. The rise and the fall time of the voltage pulse is taken to be 0.1 ns. Magnetization states visited due to the application of voltage pulse of PW=1.1 ns (top) and PW=0.5 ns (bottom) are shown to elaborate the dependence of pulse width on the final magnetization state. (NOTE: No thermal noise included here).

When the voltage pulse is withdrawn, the perpendicular anisotropy in the system is restored. To reduce the anisotropy energy, spins are forced to point in the direction perpendicular to the x-y plane, i.e. along the z-axis. Therefore, restoration of PMA makes the up or down directions (i.e. ±z directions) the preferred orientations. Depending on the pulse width (PW) of the applied voltage pulse, the magnetization can be driven to upward or downward oriented ferromagnetic state. Magnetization dynamics for five different pulse width are shown in Fig. 3. The magnetization reversal cannot take place if the pulse width is smaller than the time it takes to form a skyrmion (Fig 3, PW=0.4 ns) or if the pulse withdrawal time coincides with the expansion of skyrmion (Fig. 3, PW=1.1 ns). Successful switching can be accomplished only if the pulse withdrawal coincides with skyrmion inbreathing (i.e. shrinking) motion (Fig. 3, PW=0.5-0.7 ns). Restoration of PMA by withdrawing the voltage pulse when the skyrmion is breathing in promotes further shrinkage of the core. Therefore, the skyrmionic state annihilates and transforms into a downward ferromagnetic state.

This allows us to achieve a complete magnetization reversal employing only a unipolar voltage pulse. We note that, this method has some anology to the precessional switching scheme of ferromagnets without any DMI where the magnetization precesses around the in plane magnetic field (that we wanted to avoid in the first place). However this technique is different and potentially better in many ways:

1. No bias magnetic field is needed.

2. The skyrmion mediated reversal is oscillallatory in nature but nevertheless does not involve precession. Prior precessional schemes without DMI that involve precession around a effective magnetic field, suffer from large error rates, possibly due to incoherence in magnetization dynamics, i.e. deviation from the single domain (macrospin) state affects the switching. However, in our scheme, we accept that the reversal mechanism is inherently incoherent and then device a way to use DMI to provide a robust and reliable pathway among the multitide of pathways available for incoherent switching. This makes scheme physically different and potentially more robust to thermal noise than other precessional schemes for swicthing ferromagnets. This is corroborated by rigorously simulating the switching behavior and calculating the switching probability using various voltage pulse width in the presence of room temperature thermal noise (Fig. 4 a). An upward ferromagnetic state was relaxed for 1 ns and was subjected to voltage pulse of pulse width ranging from 0.4 ns to 1 ns. The voltage pulse is same as shown in Fig 3. Due to computational limitation only 100 events were simulated for every case. The probability shows an oscilatory behavior and between 0.46 ns to 0.58 ns all simulated events switched successfully. We picked a pulse width of 0.52 ns which is in the middle of this range and performed 10,000 successful simulations which proves the reliability of our proposed switching scheme.

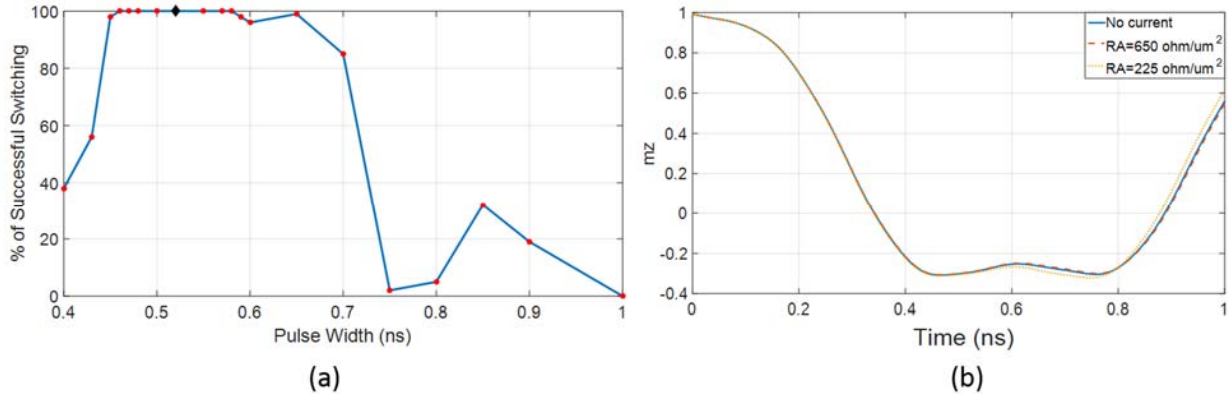

Figure 4. (a) Switching probability vs pulse width. The diamond shows the pulse width for which 10,000 simulations were carried out. The blue line is a guide to the eye. *Thermal noise is included in the magnetization dynamics.* (b) No significant variation in the magnetization dynamics due to varation in resistance area product 225-650 ohm/um$^2$ [value of RA produce based on Ref 47] which causes a small amount of perturbative spin current in addition to the VCMA (in case of perturbative current simulation, no thermal noise included).

3. Finally, with any "purely" voltage control of magnetic anisotropy (VCMA) scheme, there is always some perturbative spin current (although much smaller than STT switching) that can lead to switching error. Furthermore, across a wafer, the resistance area product can vary significantly (Ref 47 estimates the variation to be in the range of 225-650 ohm/um$^{^2}$. Our simulations (Fig. 4 b) show that the magnetization dynamics involving the skyrmion mediated switching is extremely robust to this perturbative spin current.

The modulation of the perpendicular magnetic anisotropy $\Delta$PMA = $a$E, where $a$, and E are respectively the coefficient of electric field control of magnetic anisotropy, the applied electric field. The coefficient of electric field control of magnetic anisotropy is defined as, $a = \frac{\Delta k \times t_{free}}{\Delta V / t_{MgO}}$. Reported value of "$a$" is $\approx$100 $\mu J/m^2$ per V/nm. Thus, with a 1 nm thick free layer and 1 nm thick MgO layer, $1\times10^5\ J/m^3$ change in the anisotropy energy density can be obtained per volt. The required modulation of PMA can achieved by applying a voltage pulse of 1.6 V for the proposed device configuration. These values translate into an energy dissipation of $\approx$ 0.6 fJ per switching cycle at a switching speed of ~2 GHz if all the energy required to charge the capacitive MgO layer (relative permittivity $\approx$ 7, thickness $\approx$ 1nm, and diameter $\approx$ 100 nm) is ultimately dissipated.

Previous studies showed that DMI deteriorates the thermal energy barrier of a MTJ cell[44], [45]. For the DMI value used in our simulations, the thermal energy barrier can decrease by 20%. In other words, to maintain the same thermal stability we need a $K_{eff}$ that is 20% higher than that of a system without any DMI. Additionally, one needs to reduce the anisotropy so that the skyrmion emerges as a stable state. This reduction is larger than the value needed to drive the anisotropy just in plane which is required for a system without DMI. Nevertheless, the energy dissipation is not significantly different (same order of magnitude as other VCMA schemes). As such, any non-volatile memory scheme with potential to switch a bit with < 1 fJ/bit energy dissipation is extremely competitive and 2 orders of magnitude more energy efficient than spin transfer torque random access memory (STT-RAM devices).

In summary, our proposed method, is more robust to thermal noise, perturbative spin currents and does not need a bias magnetic field, unlike VCMA induced reversal of magnetization without DMI where the magnetization rotates to upward or downward ferromagnetic state with equal probability when the voltage pulse is withdrawn and precise precessional dynamics is required for switching. The key finding in this work is that the inclusion of a small DMI in the system can create an alternative pathway for skyrmion mediated incoherent reversal between two stable single domain ferromagnetic states. Therefore, deterministic magnetization reversal is possible without depending on the precessional motion of the magnetization about an in plane magnetic field.

**Method:**

The magnetization dynamics is found by discretizing the geometry into 2×2×1 nm$^3$ cells and solving the Landau-Lifshitz-Gilbert (LLG) equation via micromagnetic simulation software-Mumax[48]

$$\frac{\partial \vec{m}}{\partial t} = \vec{\tau} = \left(\frac{-\gamma}{1+\alpha^2}\right)\left(\vec{m} \times \vec{H}_{eff} + \alpha\left(\vec{m} \times \left(\vec{m} \times \vec{H}_{eff}\right)\right)\right) \qquad (1)$$

Eqn. 1 is the LLG equation where $\vec{m}$ is the reduced magnetization ($\vec{M}/M_s$), $M_s$ is the saturation magnetization, $\gamma$ is the gyromagnetic ratio and $\alpha$ is the Gilbert damping coefficient. The effective magnetic field $\vec{H}_{eff}$ is given by,

$$\vec{H}_{eff} = \vec{H}_{demag} + \vec{H}_{exchange} + \vec{H}_{anis} + \vec{H}_{thermal} \qquad (2)$$

Here, $H_{demag}$ is the effective field due to demagnetization energy, $H_{exchange}$ is the effective field due to Heisenberg exchange coupling and DMI interaction. The DMI contribution to the effective exchange field is given by [32]:

$$H_{DM} = \frac{2D}{\mu_0 M_s}\left[(\vec{\nabla}\cdot\vec{m})\hat{z} - \vec{\nabla}m_z\right] \qquad (3)$$

where $m_z$ is the z-component of magnetization and D is the effective DMI constant.

$H_{anis}$, the effective field due to the perpendicular anisotropy, can be expressed as,

$$\vec{H}_{anis} = \frac{2K_{u1}}{\mu_0 M_{sat}}(\vec{u}\cdot\vec{m})\vec{u} + \frac{4K_{u2}}{\mu_0 M_{sat}}(\vec{u}\cdot\vec{m})^3\vec{u} \qquad (4)$$

where, $K_{u1}$ and $K_{u2}$ are first and second order uniaxial anisotropy constants and $\vec{u}$ is the unit vector in the direction of the anisotropy (i.e. perpendicular anisotropy in this case). VCMA effectively modulates the anisotropy energy density, which is given by $\Delta PMA = aE$. Here $a$ and E are respectively the coefficient of electric field control of magnetic anisotropy and the applied electric field. The resultant change in uniaxial anisotropy due to VCMA is incorporated by modulating $K_{u1}$ while keeping $K_{u2} = 0$.

We only simulated the free layer of our device. This does not affect our results as voltage induced change in the fixed layer PMA and dipolar interaction between the free and the fixed layer can both be negligible when a pinning synthetic antiferromagnetic layer is used on top of the fixed layer [49]. Additionally, in the fixed layer, the coefficient of electric field control of magnetic anisotropy can be tailored to be low and PMA can be designed to be strong to further ensure the magnetization direction of fixed layer is not perturbed due to application of a voltage pulse.

Recent studies reported modification of DMI and exchange stiffness by applying Electric field [50], [51] which are not considered in our model. However, increase (decrease) in the exchange stiffness due to a positive (negative) electric field will assist skyrmion-ferromagnetic (ferromagnetic-skyrmion) transformation. We verified this by simulating scenarios considering electric field induced modification of exchange stiffness and found that switching occurs at lower $\Delta$PMA. Hence, the estimated voltage or energy dissipation in this study can be considered conservative. On the other hand, DMI modulation due to electric field is minimal and therefore, was ignored.